\documentclass[12pt]{article}
\pdfoutput=1

\usepackage{graphics,amssymb,float}
\usepackage{graphicx}
\usepackage{rotating}
\usepackage{dcolumn}
\usepackage{bm}
\usepackage{cite}
\usepackage{amsmath}
\usepackage{indentfirst} 
\usepackage{epstopdf}
\usepackage[usenames,dvipsnames]{xcolor}

\makeatletter

\@addtoreset{equation}{section}
\makeatother

\def\lsim{\;\raise0.3ex\hbox{$<$\kern-0.75em\raise-1.1ex\hbox{$\sim$}}\;}
\def\gsim{\;\raise0.3ex\hbox{$>$\kern-0.75em\raise-1.1ex\hbox{$\sim$}}\;}
\def\beq{\begin{equation}}   \def\eeq{\end{equation}}
\def\ba{\begin{array}}       \def\ea{\end{array}}
\def\bea{\begin{eqnarray}}   \def\eea{\end{eqnarray}}
\def\nn{\nonumber}
\def\nl{\newline}

\begin{document}

\begin{titlepage}
\begin{flushright}
LPT Orsay 14-30 \\
LUPM 14-015\\
\end{flushright}


\begin{center}
\vspace{1cm}
{\Large\bf The semi-constrained NMSSM satisfying bounds from the LHC,
LUX and Planck} \\
\vspace{2cm}

{\bf{Ulrich Ellwanger$^a$ and Cyril
Hugonie$^b$}}\\
\vspace{1cm}
\it  $^a$ LPT, UMR 8627, CNRS, Universit\'e de Paris--Sud, 91405 Orsay,
France \\
\it $^b$ LUPM, UMR 5299, CNRS, Universit\'e de Montpellier II, 34095
Montpellier, France \\

\end{center}
\vspace{2cm}

\begin{abstract}
We study the parameter space of the semi-constrained NMSSM, compatible
with constraints on the Standard Model like Higgs mass and signal
rates, constraints from searches for squarks and gluinos,
a dark matter relic density compatible with bounds from WMAP/Planck, and
direct detection cross sections compatible with constraints from LUX.
The remaining parameter space allows for a fine-tuning as low as about
100, an additional
lighter Higgs boson in the 60-120~GeV mass range detectable in the
diphoton mode or in decays into a pair of lighter CP-odd Higgs
bosons, and dominantly singlino like dark matter
with a mass down to 1~GeV, but possibly a very small direct detection
cross section.
\end{abstract}

\end{titlepage}

\section{Introduction}

Recent results from the LHC and direct dark matter detection experiments
constrain considerably possible scenarios beyond the Standard Model
(SM), amongst others its supersymmetric (SUSY) extensions. These
constraints originate essentially from the Higgs mass~\cite{Aad:2012tfa,
Chatrchyan:2012ufa} and its quite SM like signal rates, the absence of
signals in searches for squarks and gluinos after the 8~TeV run at the
LHC~\cite{TheATLAScollaboration:2013fha, Chatrchyan:2014lfa}, and
upper bounds on dark matter--nucleus cross sections from the LUX
experiment~\cite{Akerib:2013tjd}.

Masses and couplings of Higgs boson(s), SUSY particles and notably the
lightest SUSY particle (LSP, the dark matter candidate), are strongly
correlated in SUSY extensions of the SM if one assumes at least
partial unification of the soft SUSY breaking terms at a grand
unification (GUT) scale.
Hence it is interesting to study how the combined constraints affect the
parameter space and, notably, which signals beyond the SM we can expect
in the future. Such studies (after the discovery of the 126~GeV Higgs
boson) had been performed earlier in the Minimal SUSY extension of the
SM (MSSM)~\cite{ Baer:2011ab, Arbey:2011ab, Buchmueller:2011ab,
Akula:2011aa, Kadastik:2011aa, Cao:2011sn, Ellis:2012aa, Fowlie:2012im,
Beskidt:2012sk, Buchmueller:2012hv, Strege:2012bt, Ellis:2012nv,
Cabrera:2012vu, Kowalska:2013hha, Cohen:2013kna, Beskidt:2013gia,
Henrot-Versille:2013yma, Bechtle:2013mda, Kim:2013uxa,
Buchmueller:2013rsa, Ellis:2013oxa} and the Next-to-Minimal SUSY
extension of the SM (NMSSM)~\cite{Arbey:2011ab, Gunion:2012zd,
Ellwanger:2012ke, Belanger:2012tt, Kowalska:2012gs,
Kim:2013uxa,Beskidt:2014oea}.

These studies differ, however, in the treatment of the soft SUSY
breaking terms in the Higgs sector at the GUT scale: In ``fully
constrained'' versions of the MSSM or NMSSM these are supposed to be
unified with the soft SUSY breaking terms in the squark and slepton
sectors. In NUHM (non-universal Higgs masses) or ``semi-constrained''
versions of the MSSM or NMSSM one allows the soft SUSY breaking terms in
the Higgs sector to be different; after all the quantum numbers of the
Higgs fields differ from those of quarks and leptons: Higgs fields are
in a real representation ($2 + \bar{2}$) of SU(2), but do not fit into
complete representations of SU(5); these properties can easily have an
impact on the presently unknown sources of soft SUSY breaking terms. In
the NMSSM including the singlet superfield $S$, ``semi-constrained'' can
indicate non-universal soft SUSY breaking terms involving the singlet
only, or non-universal soft SUSY breaking terms involving SU(2) doublet
or singlet Higgs fields. In the present study we allow for the latter
more general case.

Previous studies of the NMSSM with constraints at the GUT scale~\cite{
Gunion:2012zd, Ellwanger:2012ke, Belanger:2012tt, Kowalska:2012gs,
Kim:2013uxa, Beskidt:2014oea} had found that wide ranges of parameter
space comply with constraints from the LHC and on dark matter, and that
less ``tuning'' is required than in the MSSM~\cite{Kowalska:2012gs,
Kim:2013uxa}. These findings are confirmed by scans of the parameter
space of the general NMSSM (without constraints at the GUT scale)~\cite{
Kang:2012sy, Cao:2012fz, Vasquez:2012hn, Perelstein:2012qg,Agashe:2012zq,
Gherghetta:2012gb, Cheng:2013fma}, and motivate a thorough analysis of
the semi-constrained NUH-NMSSM with up-to-date experimental constraints,
amongst others on Higgs signal rates and bounds on dark matter--nucleus
cross sections~\cite{Akerib:2013tjd}. ``NUH'' appears without ``M''
since, apart from the Higgs mass terms, also trilinear couplings
involving Higgs bosons only are allowed to differ from trilinear
couplings involving squarks or sleptons at the GUT scale, see the next
section.

Using the code {\sf NMSPEC}~\cite{Ellwanger:2006rn} within {\sf
NMSSMTools\_4.2.1}~\cite{Ellwanger:2004xm,Ellwanger:2005dv}
together with {\sf micr}\-{\sf OMEGAS\_3} \cite{Belanger:2013oya} we
have sampled about 3.2~M viable points in the
parameter space, which allows us to cover the complete range of masses
and couplings of the LSP and additional Higgs bosons, parts of which had
not been observed in previous analyses. In this paper we confine
ourselves to regions where an additional NMSSM-specific Higgs scalar is
lighter than the SM-like Higgs boson near 126~GeV; this region is
strongly favoured by the mass of the SM-like Higgs boson, and contains
the most interesting phenomena to be searched for in the future.

In the next section we present the model, the applied phenomenological
constraints, the definition of fine-tuning, and the ranges of
parameters scanned over. In section~3 we discuss the impact of
unsuccessful searches for squarks and gluinos at the LHC on fine-tuning
and some of the parameters like the soft squark/slepton masses $m_0$,
the universal gaugino masses $M_{1/2}$ and the NMSSM-specific Yukawa
coupling $\lambda$. Section~4 is devoted to the properties of the LSP,
its detection rates to be expected in the future, and its annihilation
processes allowing for a viable relic density. In section~5 we discuss
the Higgs sector, in particular prospects to detect the lighter NMSSM
specific Higgs scalar. Conclusions and an outlook are given in
section~6.

\section{The NMSSM with constraints at the GUT scale}

The NMSSM~\cite{Ellwanger:2009dp} differs from the MSSM due to the
presence of the gauge singlet superfield $S$. In the simplest $\mathbb
Z_3$ invariant realisation of the NMSSM, the Higgs mass term $\mu H_u
H_d$ in the superpotential $W_\mathrm{MSSM}$ of the MSSM is replaced by
the coupling $\lambda$ of $S$ to $H_u$ and $H_d$ and a self-coupling
$\kappa S^3$.  Hence, in this simplest version the superpotential
$W_\mathrm{NMSSM}$ is scale invariant and given by
\beq\label{eq:2.1}
W_\mathrm{NMSSM} = \lambda \hat S \hat H_u\cdot \hat H_d +
\frac{\kappa}{3} 
\hat S^3 + \dots\; ,
\eeq
where hatted letters denote superfields, and the ellipses denote the
MSSM-like Yukawa couplings of $\hat H_u$ and $\hat H_d$ to the
quark and lepton superfields. Once the real scalar component of
$\hat S$ develops a vev $s$, the first term in $W_\mathrm{NMSSM}$
generates an effective $\mu$-term
\beq\label{eq:2.2}
\mu_\mathrm{eff}=\lambda\, s\; .
\eeq

The soft Susy breaking terms consist of mass terms for the Higgs bosons
$H_u$, $H_d$, $S$, squarks
$\tilde{q_i} \equiv (\tilde{u_i}_L, \tilde{d_i}_L$), $\tilde{u_i}_R^c$,
$\tilde{d_i}_R^c$ and sleptons $\tilde{\ell_i} \equiv (\tilde{\nu_i}_L,
\tilde{e_i}_L$) and $\tilde{e_i}_R^c$ 
(where $i=1,2,3$ is a generation index):
\bea
-{\cal L}_\mathrm{0} &=&
m_{H_u}^2 | H_u |^2 + m_{H_d}^2 | H_d |^2 + 
m_{S}^2 | S |^2 +m_{\tilde{q_i}}^2|\tilde{q_i}|^2 
+ m_{\tilde{u_i}}^2|\tilde{u_i}_R^c|^2
+m_{\tilde{d_i}}^2|\tilde{d_i}_R^c|^2\nn \\
&& +m_{\tilde{\ell_i}}^2|\tilde{\ell_i}|^2
+m_{\tilde{e_i}}^2|\tilde{e_i}_R^c|^2\; ,
\label{eq:2.3}
\eea
trilinear interactions involving the third generation squarks, sleptons and the
Higgs fields (neglecting the Yukawa couplings of the two first generations):
\bea
-{\cal L}_\mathrm{3}&=& 
\Bigl( h_t A_t\, Q\cdot H_u \: \tilde{u_3}_R^c +
h_b  A_b\, H_d \cdot Q \: \tilde{d_3}_R^c +
h_\tau A_\tau \,H_d\cdot L \: \tilde{e_3}_R^c  \nn \\
&& +\,  \lambda A_\lambda\, H_u  \cdot H_d \,S +  \frac{1}{3} \kappa 
A_\kappa\,  S^3 \Bigl)+ \, \mathrm{h.c.}\;,
\label{eq:2.4}
\eea
and mass terms for the gauginos $\tilde{B}$ (bino), $\tilde{W}^a$
(winos) and $\tilde{G}^a$ (gluinos):
 \beq\label{eq:2.5}
-{\cal L}_\mathrm{1/2}= \frac{1}{2} \bigg[ 
 M_1 \tilde{B}  \tilde{B}
\!+\!M_2 \sum_{a=1}^3 \tilde{W}^a \tilde{W}_a 
\!+\!M_3 \sum_{a=1}^8 \tilde{G}^a \tilde{G}_a   
\bigg]+ \mathrm{h.c.}\;.
\eeq

In constrained versions of the NMSSM one assumes that the soft Susy
breaking terms involving gauginos, squarks or sleptons are identical at
the GUT scale:
\beq \label{eq:2.6}
M_1 = M_2 = M_3 \equiv M_{1/2}\; ,
\eeq
\beq \label{eq:2.7}
m_{\tilde{q_i}}^2= m_{\tilde{u_i}}^2=m_{\tilde{d_i}}^2=
m_{\tilde{\ell_i}}^2=m_{\tilde{e_i}}^2\equiv m_0^2\; ,
\eeq
\beq \label{eq:2.8}
A_t = A_b = A_\tau \equiv A_0\; .
\eeq

In the NUH-NMSSM considered here one allows the Higgs sector to play a
special role: The Higgs soft mass terms $m_{H_u}^2$, $m_{H_d}^2$ and
$m_{S}^2$ are allowed to differ from $m_0^2$ (and determined implicitely
at the weak scale by the three minimization equations of the effective potential),
and the trilinear
couplings $A_{\lambda}$, $A_{\kappa}$ can differ from $A_0$. Hence the
complete parameter space is characterized by
\beq \label{eq:2.9}
\lambda\ , \ \kappa\ , \ \tan\beta\ ,\ \mu_\mathrm{eff}\ , \ A_{\lambda}
\ , \ A_{\kappa} \ , \ A_0 \ , \ M_{1/2}\ ,  \ m_0\; ,
\eeq
where the latter five parameters are taken at the GUT scale.

Expressions for the mass matrices of the physical CP-even and CP-odd
Higgs states -- after $H_u$, $H_d$ and $S$ have assumed vevs $v_u$,
$v_d$ and $s$ and including the dominant radiative corrections -- can be
found in~\cite{Ellwanger:2009dp} and will not be repeated here. The
physical CP-even Higgs states will be denoted as $H_i$, $i=1,2,3$
(ordered in mass), and the physical CP-odd Higgs states as $A_i$,
$i=1,2$. The neutralinos are denoted as $\chi^0_i$, $i=1,...,5$ and
their mixing angles $N_{i,j}$ such that $N_{1,5}$ indicates the singlino
component of the lightest neutralino $\chi^0_1$.

Subsequently we are interested in regions of the parameter space where
doublet-singlet mixing in the Higgs sector leads to an increase of the
mass of the SM-like (mostly doublet-like) Higgs boson, which leads
naturally to a SM-like Higgs boson $H_2$ in the 126~GeV range~\cite{
Hall:2011aa,Ellwanger:2011aa,Arvanitaki:2011ck,King:2012is, Kang:2012sy,
Cao:2012fz}, but implies a lighter mostly singlet-like Higgs state
$H_1$.

Recent phenomenological constraints include, amongst others, upper
bounds on the direct (spin independent) detection rate of dark matter by
LUX~\cite{Akerib:2013tjd}. In the NMSSM, the LSP (the dark matter candidate)
is assumed to be the lightest neutralino, as in the MSSM.
Its spin independent detection rate and relic
density are computed with the help of
{\sf micrOMEGAS\_3} ~\cite{Belanger:2013oya}.
We apply the upper bounds of LUX and
require a relic density inside a slightly enlarged WMAP/Planck
window~\cite{Hinshaw:2012aka,Ade:2013zuv} $0.107 \leq \Omega h^2 \leq
0.131$ in order not to loose too many points in parameter space; the
precise value of $\Omega h^2$ has little impact on the subsequent
results.

In the Higgs sector we require a neutral CP-even state with a mass of
$125.7 \pm 3$~GeV allowing for theoretical and parametric uncertainties
of the mass calculation; we used 173.1~GeV for the top quark mass. Its
signal rates should comply with the essentially SM-like signal rates in
the channels measured by ATLAS/CMS/Teva\-tron. These measurements can be
combined leading to 95\% confidence level (CL) contours in the planes of
Higgs production via (gluon fusion and ttH) -- (vector boson fusion and 
associate
production with W/Z), separately for Higgs decays into $\gamma\gamma$,
ZZ or WW and $b\bar{b}$ or  $\tau^+\tau^-$. We require that the signal
rates for a Higgs boson in the above mass range are within all three
95\% confidence level contours derived in~\cite{Belanger:2013xza}.

The application of constraints from unsuccessful searches for sparticles
at the first run of the LHC is more delicate: These bounds depend on all
parameters of the model via the masses and couplings (and the resulting
decay cascades) of all sparticles. However, it is possible to proceed as
follows, using the most constraining searches for gluinos and squarks of
the first generation in events with jets and missing $E_T$: For heavy
squarks and/or gluinos the production cross sections are so small that
these points in parameter space are not excluded independently of the
squark/gluino decay cascades. On the other hand, relatively light
squarks and/or gluinos are excluded independently of their decay
cascades. In between these regions defined in the planes of
squark/gluino masses or $m_0/M_{1/2}$, exclusion does depend on their
decays, in particular on the presence of a light singlino-like LSP at
the end of the cascades~\cite{Das:2012rr,Das:2013ta}. 

The boundaries between these three regions were obtained with
the help of the analysis of some hundreds of points in parameter space:
Events were generated by MadGraph/MadEvent \cite{Alwall:2011uj} which
includes Pythia~6.4~\cite{Sjostrand:2006za} for showering and
hadronisation. The sparticle branching ratios are obtained with the help
of the code NMSDECAY~\cite{Das:2011dg} (based on
SDECAY~\cite{Muhlleitner:2003vg}), and are passed to Pythia. The output in
StdHEP format is given to CheckMATE~\cite{Drees:2013wra} which includes
the detector simulation DELPHES~\cite{deFavereau:2013fsa} and compares
the signal rates to constraints in various search channels of ATLAS and
CMS. Corresponding results will be presented in section~3.

Other constraints from $b$-physics, LEP (from Higgs searches and
invisible Z decays) and the LHC (on heavy Higgs bosons decaying into
$\tau^+ \tau^-$) are applied as in
{\sf NMSSMTools\_4.2.1}~\cite{Ellwanger:2004xm,Ellwanger:2005dv},
leaving aside the muon anomalous magnetic moment.

Since the fundamental parameters of the model are the masses and
couplings at the GUT scale, it makes sense to ask in how far these have
to be tuned relative to each other in order to comply with the SM-like
Higgs mass and the non-observation of sparticles at the LHC. To this end
we consider the usual measure of fine-tuning~\cite{Barbieri:1987fn}
\beq\label{eq:2.10}
FT = Max\left\{\left|\frac{\partial \ln(M_Z)}{\partial \ln(p_i^{GUT})}
\right|\right\}
\eeq
where $p_i^{GUT}$ denote all dimensionful and dimensionless parameters
(Yukawa couplings, mass terms and trilinear couplings) at the GUT scale.
$FT$ is computed numerically in {\sf NMSSMTools\_4.2.1} following the
method described in~\cite{Ellwanger:2011mu} where details can be found.

We have scanned the parameter space of the NUH-NMSSM given
in~(\ref{eq:2.9}) using a Markov Chain Monte Carlo (MCMC) technique. In
addition to the phenomenological constraints discussed above we require
the absence of Landau singularities of the running Yukawa couplings
below the GUT scale, and the absence of deeper unphysical minima of the
Higgs potential with at least one vanishing vev $v_u$, $v_d$ or $s$.
Bounds on the dimensionful parameters follow from
the absence of too large fine-tuning; we imposed $FT < 1000$. Finally we
obtained $\sim 3.2\times10^6$ valid points in parameter space within the
following ranges of the parameters~(\ref{eq:2.9}):
\bea
1\times10^{-6} \leq \lambda \leq 0.722,& -0.08 \leq \kappa \leq 0.475,&
1.42 \leq  \tan\beta  \leq 60.3,\nn\\ 
-537\  \mathrm{GeV} \leq \mu_\mathrm{eff} \leq 753\ \mathrm{GeV},&
-19\ \mathrm{TeV} \leq A_{\lambda} \leq 8.5\ \mathrm{TeV},
&-1.3\ \mathrm{TeV} \leq A_{\kappa} \leq 5.3\ \mathrm{TeV},\nn\\
0\  \leq m_0 \leq 4.4\ \mathrm{TeV},& 0.1\ \mathrm{TeV}
 \leq M_{1/2} \leq 3.1\ \mathrm{TeV},&
 -6.6\ \mathrm{TeV} \leq A_{0} \leq 8.1\ \mathrm{TeV}.\nn\\
 &&\label{eq:2.11}
\eea
The fact that the upper bounds on the dimensionful parameters are
distinct originates from the different impact of these parameters on the
fine-tuning, which is often dominated by the universal gaugino mass
parameter $M_{1/2}$.

\section{Impact of LHC constraints on squark/gluino masses and
fine-tuning}

Strong constraints on parameter spaces of SUSY extensions of the SM come
from searches for gluinos $\tilde{g}$ and squarks $\tilde{q}$ of the
first generation in events with jets and missing
$E_T$~\cite{TheATLAScollaboration:2013fha, Chatrchyan:2014lfa}.
In~\cite{TheATLAScollaboration:2013fha} exclusion limits for MSUGRA/CMSSM
models have been given in the $m_0-M_{1/2}$ and
$M_{\tilde{g}}-m_{\tilde{q}}$ planes for $\tan\beta=30$, $A_0=-2m_0$ and
$\mu>0$.

As a result of the simulations described in the previous section we
found that the 95\%~CL upper limits on signal events
in~\cite{TheATLAScollaboration:2013fha} lead to exclusion limits in the
$m_0-M_{1/2}$ or $M_{\tilde{g}}-m_{\tilde{q}}$ planes in the NUH-NMSSM
which are very similar to the CMSSM if the LSP is bino-like, but can be
alleviated in the presence of a light singlino-like LSP at the end of
the cascades~\cite{Das:2012rr,Das:2013ta}. Still, even with a
singlino-like LSP, certain regions in these planes are always excluded.

\begin{figure}[ht!]\centering
\includegraphics[width=.5\textwidth]{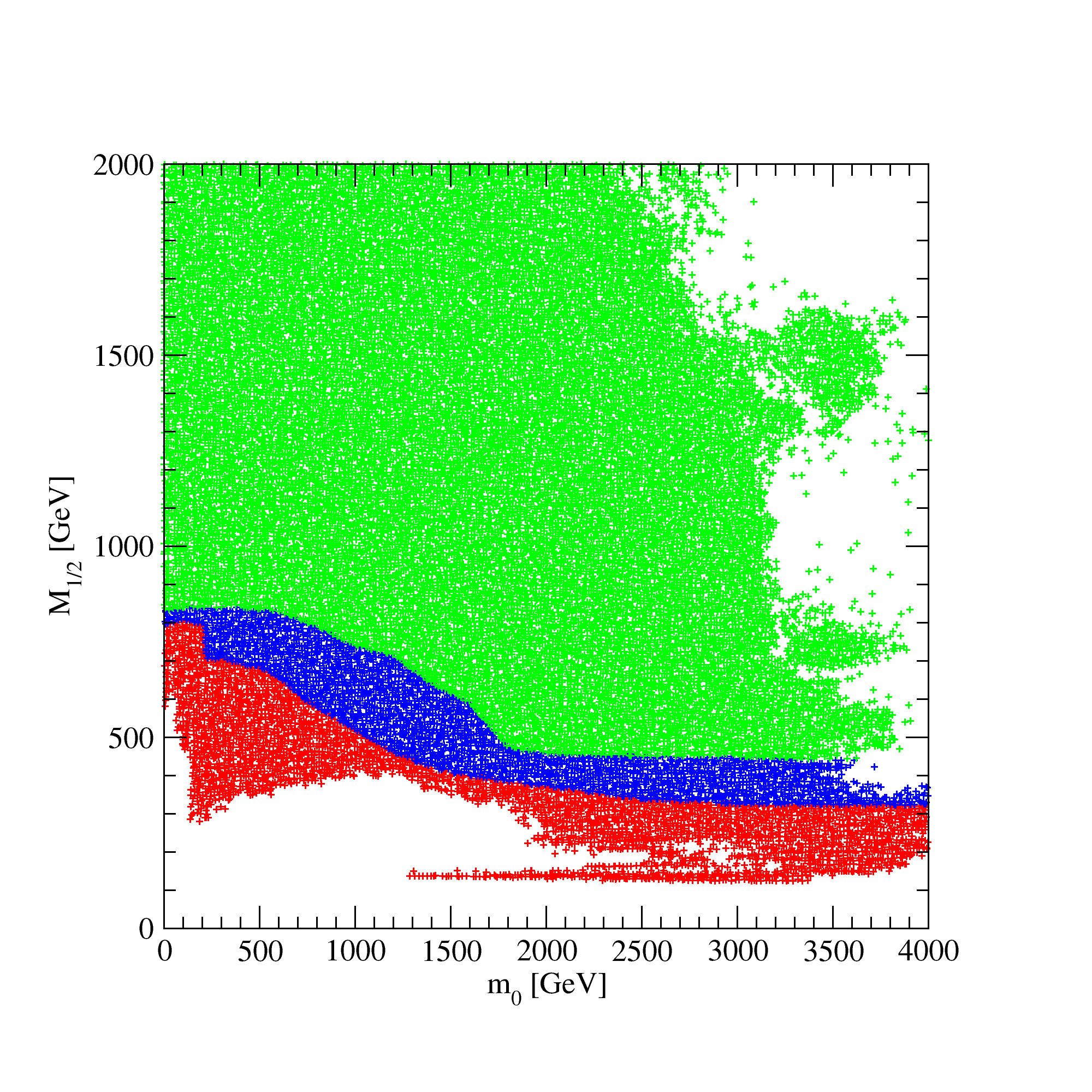}
\hspace*{-2em}
\includegraphics[width=.5\textwidth]{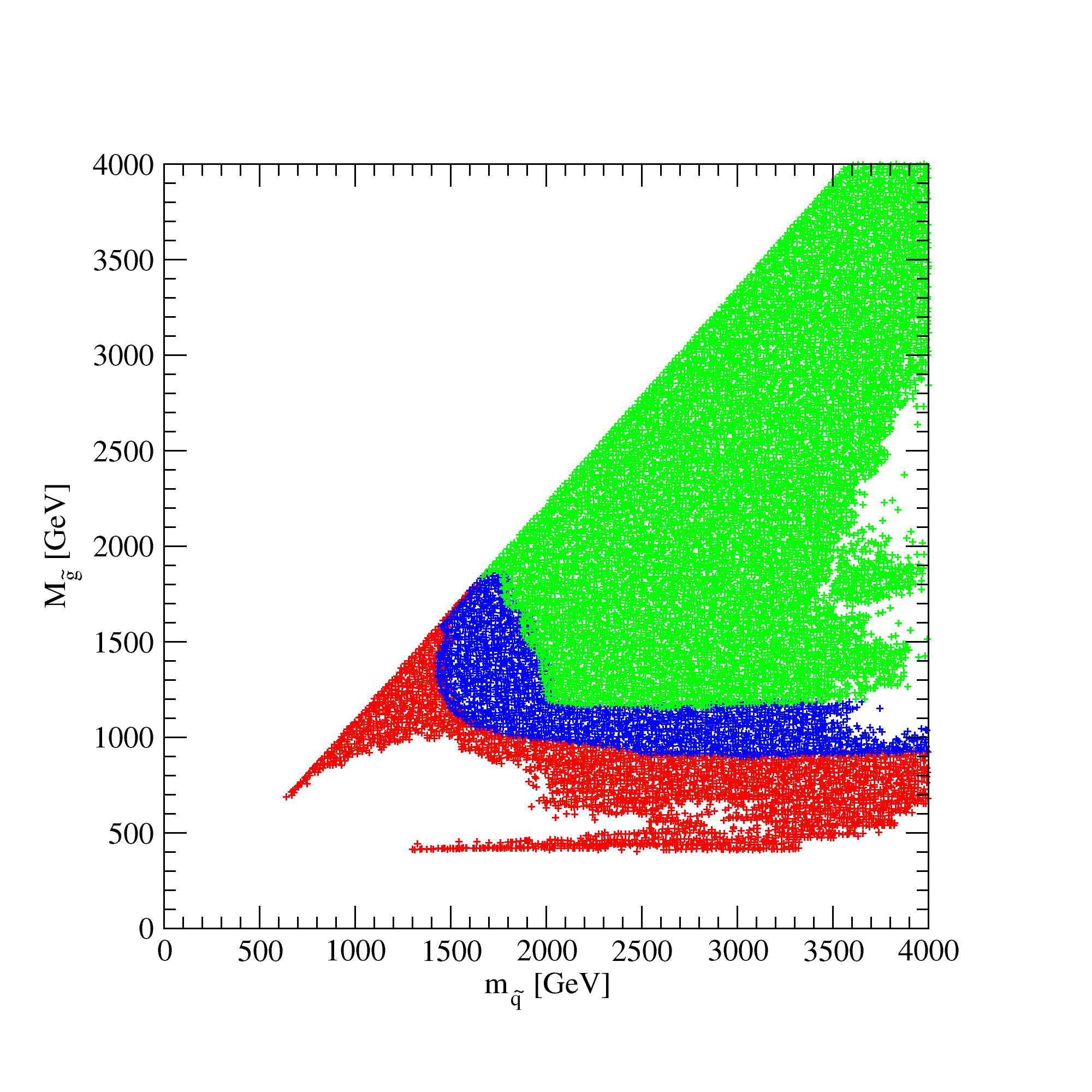}
\vspace*{-2em}
\caption{The $m_0-M_{1/2}$ and
$M_{\tilde{g}}-m_{\tilde{q}}$ planes in the NUH-NMSSM.
Green: regions allowed by the 95\% CL upper limits on signal events
in~\cite{TheATLAScollaboration:2013fha}, blue: regions allowed in
the presence of a singlino-like LSP, red: regions which are
always excluded.}
\label{fig:1}
\end{figure}

In Fig. \ref{fig:1} we show the $m_0-M_{1/2}$ and
$M_{\tilde{g}}-m_{\tilde{q}}$ planes in the NUH-NMSSM and indicate in
green the regions allowed by the 95\% CL upper limits on signal events
(practically identical to the ones given
in~\cite{TheATLAScollaboration:2013fha}), in blue the regions possibly
allowed in the presence of a singlino-like LSP, and in red the regions
which are always excluded. Note that, in contrast to the MSSM, the limit
$m_0 \to 0$ is always possible for all $M_{1/2}$: In the MSSM this
region is limited by the appearance of a stau LSP. In the NMSSM a
singlino-like LSP can always be lighter than the lightest stau, and its
relic density can be reduced to the WMAP/Planck window through
singlino-stau coannihilation as in the fully constrained
NMSSM~\cite{Djouadi:2008yj,Djouadi:2008uj} or through narrow resonances
implying specific NMSSM light Higgs states~\cite{Belanger:2005kh,
Hugonie:2007vd, Belanger:2008nt, Vasquez:2012hn}. (The combined
constraints from the Higgs sector and the nature of the LSP lead to
discontinuities in the allowed parameter space for small $m_0$.)

\begin{figure}[ht!]\centering
\includegraphics[width=.5\textwidth]{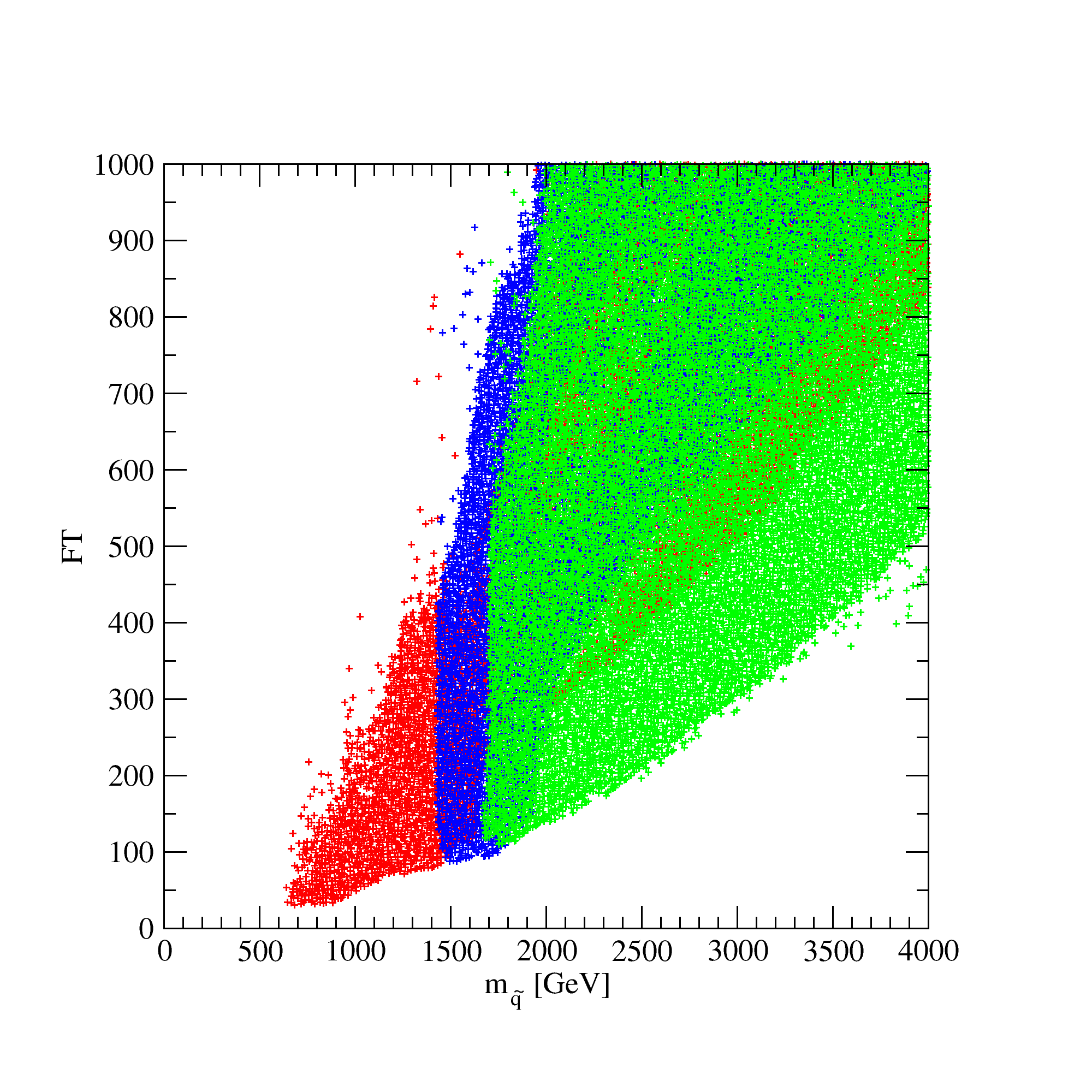}
\hspace*{-2em}
\includegraphics[width=.5\textwidth]{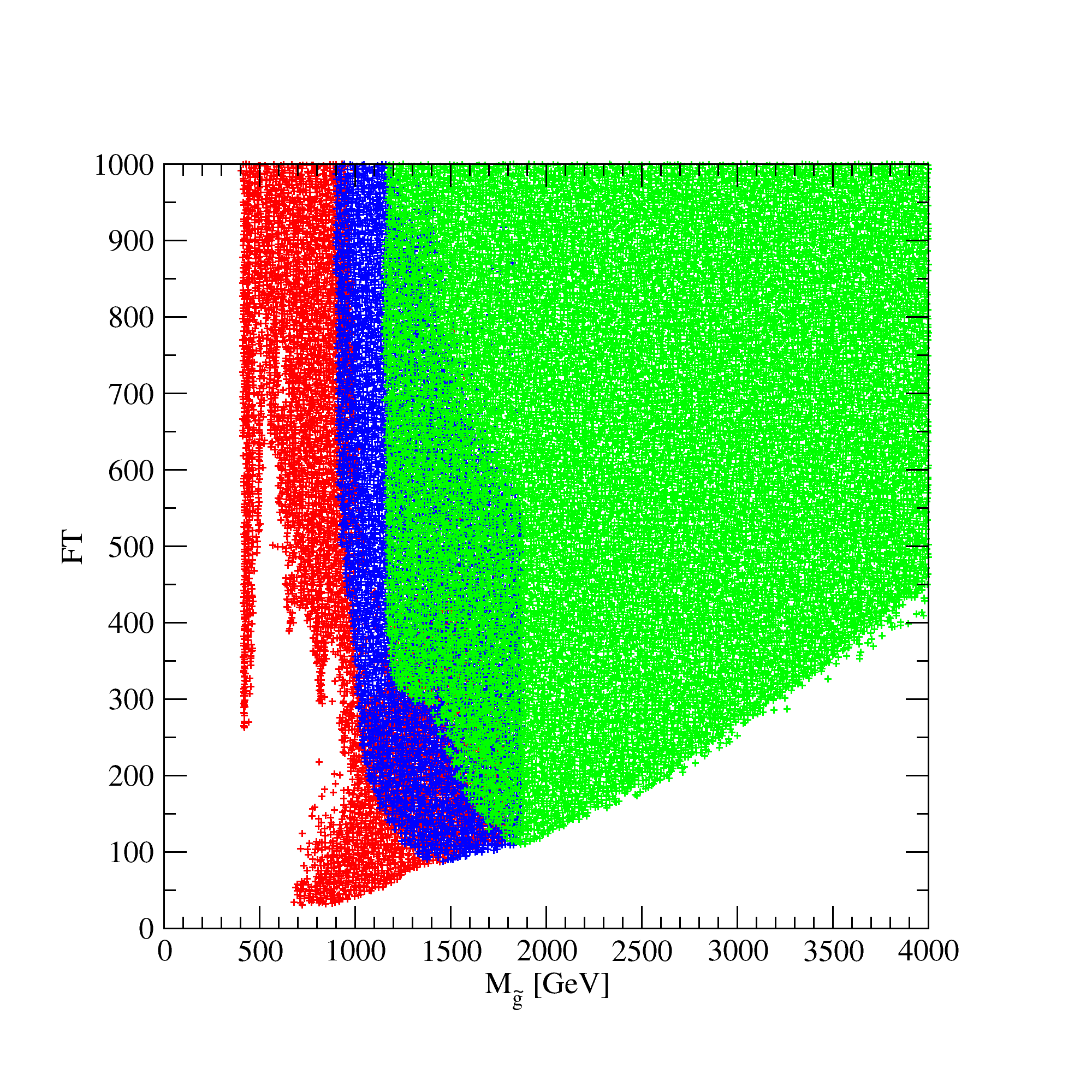}
\vspace*{-2em}
\caption{$FT$ as defined in (\ref{eq:2.10}) as function of the squark
and gluino masses, and the impact of the LHC constraints in the same
color coding as in Fig. \ref{fig:1}.}
\label{fig:2}
\end{figure}

These lower bounds on the squark and gluino masses dominate the lower
bounds on the fine-tuning $FT$ defined in (\ref{eq:2.10}). In
Fig.~\ref{fig:2} we show $FT$ as function of the squark and gluino
masses, and the impact of the LHC constraints in the same color coding
as in Fig.~\ref{fig:1}.

We see that the LHC forbidden red region increases the lower bound on
$FT$ from $\sim 20$ to $FT \gsim 80$; the NMSSM-specific alleviation
(blue region) has a minor impact on $FT$. The dominant contribution to
$FT$ in (\ref{eq:2.10}) originates typically from $M_{1/2}$ (i.e. the
gluino mass at the GUT scale), or from the soft Higgs mass term
$m_{H_u}^2$. If one requires unification of $m_{H_u}$ and $m_{H_d}$ with
$m_0$ as in~\cite{Kowalska:2012gs}, $FT$ is considerably larger $(\gsim
400)$. In the MSSM -- after imposing LHC constraints on squark and
gluino masses, defining $FT$ with respect to parameters at the GUT scale
and allowing for non-universal Higgs mass terms at the GUT scale as
in~\cite{Baer:2012cf} -- one finds $FT \gsim 1000$. The much lower value of
$FT$ in the NUH-NMSSM coincides with the result in~\cite{Binjonaid:2014oga}.

\begin{figure}[ht!]\centering
\includegraphics[width=.5\textwidth]{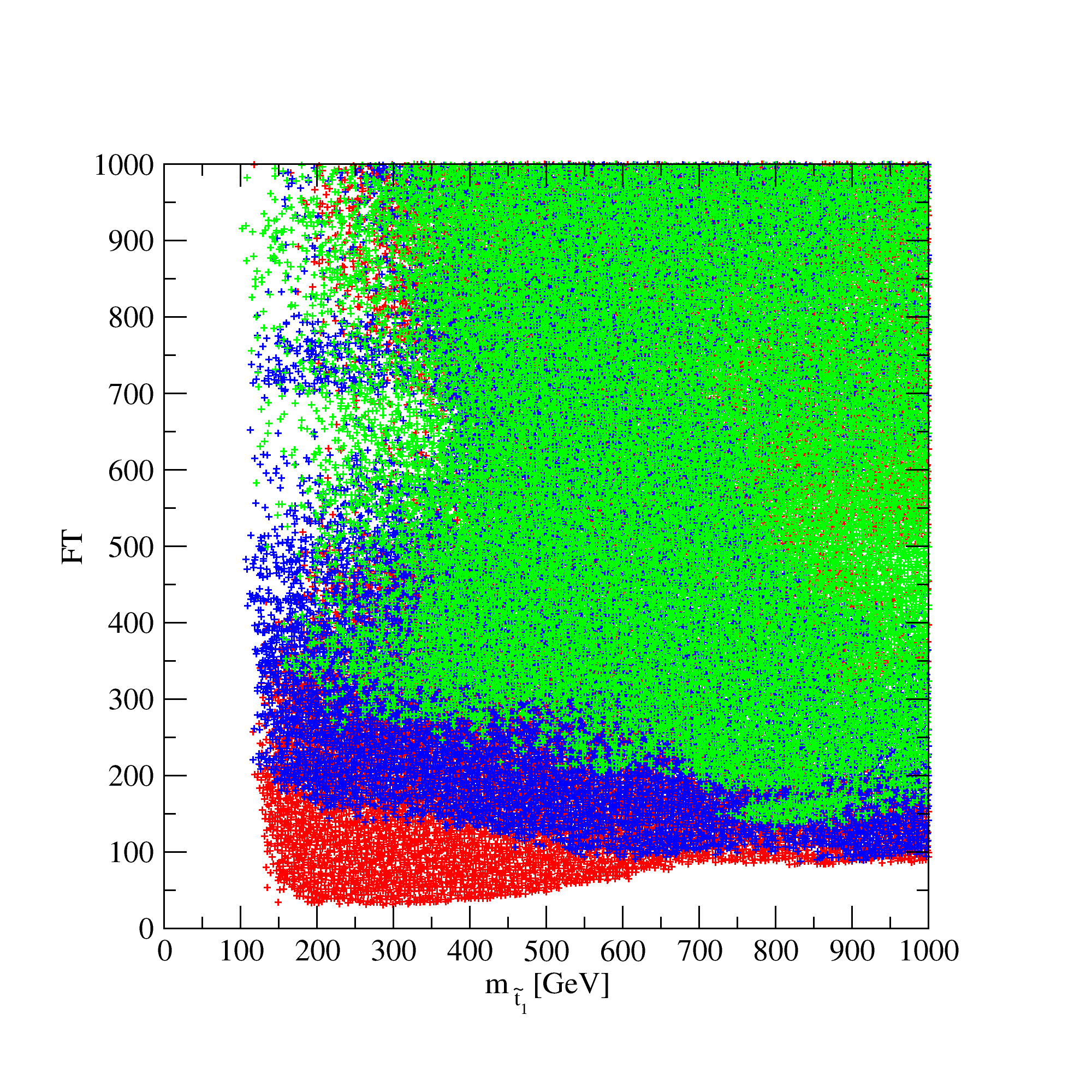}
\hspace*{-2em}
\includegraphics[width=.5\textwidth]{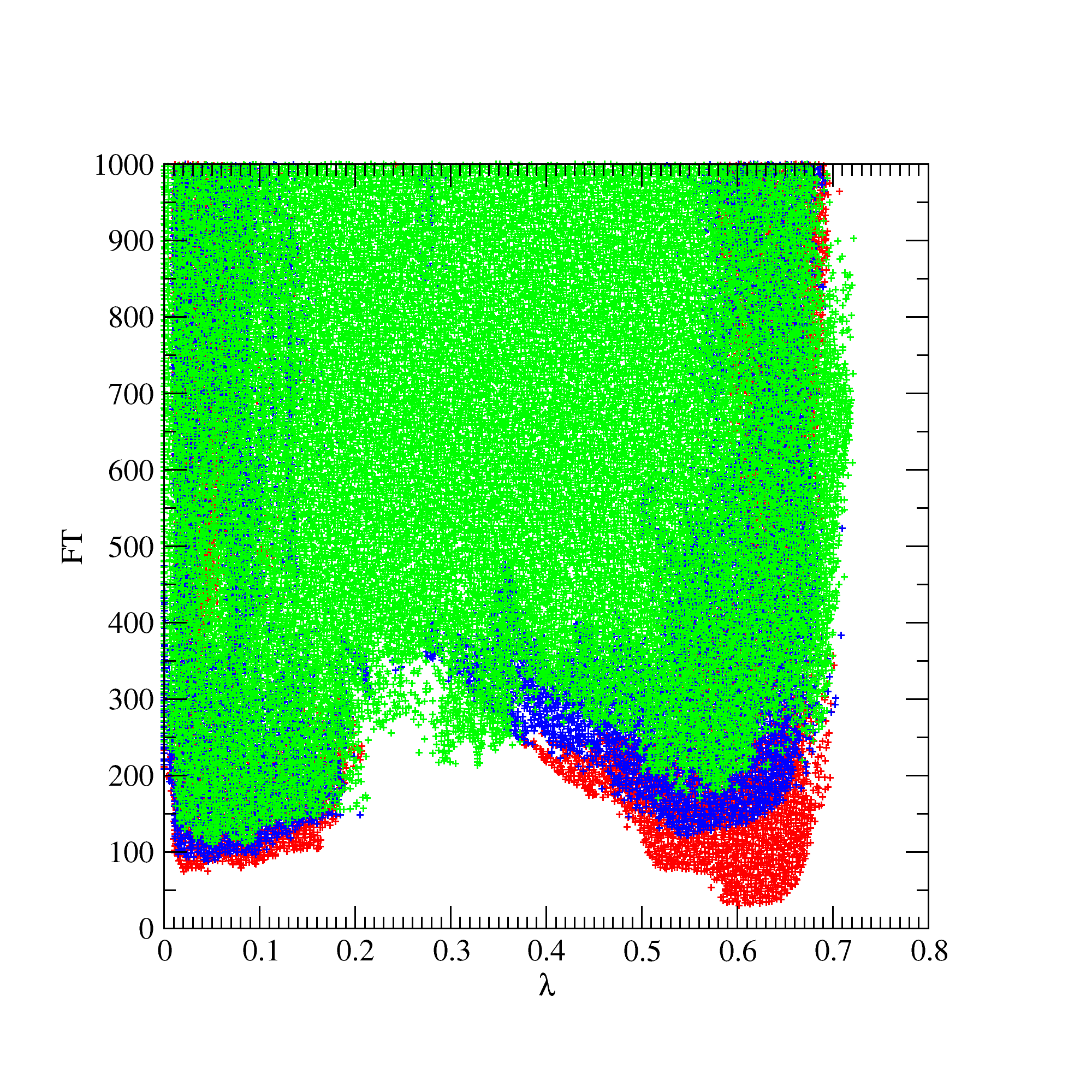}
\vspace*{-2em}
\caption{Left: $FT$ as function of the $m_{\tilde{t}_1}$. Right: $FT$ as
function of $\lambda$ at the SUSY scale. The
color coding is as in Fig.~\ref{fig:1}.}
\label{fig:3}
\end{figure}

The impact of $M_{1/2}$ on $FT$ is actually indirect: Heavy gluinos lead
to large radiative corrections to the stop masses which,
in turn, lead to large radiative corrections to the soft Higgs mass
terms. Therefore, if one defines $FT$ with respect to parameters at a
lower scale, low $FT$ is typically related to light stops. On the
left-hand side of Fig.~\ref{fig:3} we show $FT$ as function of the mass
$m_{\tilde{t}_1}$ of the lightest stop. We see that, without imposing LHC
constraints on squark and gluino masses, the lower bound on $FT$ (still
with respect to parameters at the GUT scale) would increase slightly with
$m_{\tilde{t}_1}$, but with LHC constraints the lower bound on $FT$
depends weakly on (decreases only slightly with) $m_{\tilde{t}_1}$.

In the MSSM, the measured mass of the SM-like Higgs $H_{SM}$ requires relatively
heavy stops and/or a Higgs-stop trilinear coupling $A_t$, which also
contribute to $FT$. In the NMSSM (recall that, in the scenario considered
here, $H_{SM}=H_2$) large radiative corrections to the
SM-like Higgs mass $m_{H_{SM}}$ are not required, since the SM-like
Higgs mass can be pushed upwards either through a positive tree level
contribution $\sim \lambda^2 \sin^2 2\beta$~\cite{Ellwanger:2009dp}, or
through mixing with a lighter Higgs state $H_1$~\cite{Ellwanger:1999ji}
which does not require large values of $\lambda$~\cite{Badziak:2013bda}.
(In the latter scenario too
large values of $\lambda$, i.e. a too large $H_1 - H_{SM}$ mixing angle,
can imply an inacceptable reduction of the signal rates of $H_{SM}$ at
the LHC and/or lead to the violation of LEP constraints on $H_1$.)

On the right-hand side of Fig.~\ref{fig:3} we show $FT$ as function of
$\lambda$. We see that -- without imposing LHC constraints on squark and
gluino masses -- the minimum of $FT$ would indeed be assumed for
$\lambda \sim 0.6$ related to the tree level contribution $\sim
\lambda^2 \sin^2 2\beta$ to $m_{H_{SM}}$. Including LHC constraints,
local minima of $FT$ exist both for $\lambda \approx 0.6$ and $\lambda
\approx 0.1$.

\begin{figure}[ht!]\centering
\includegraphics[width=.5\textwidth]{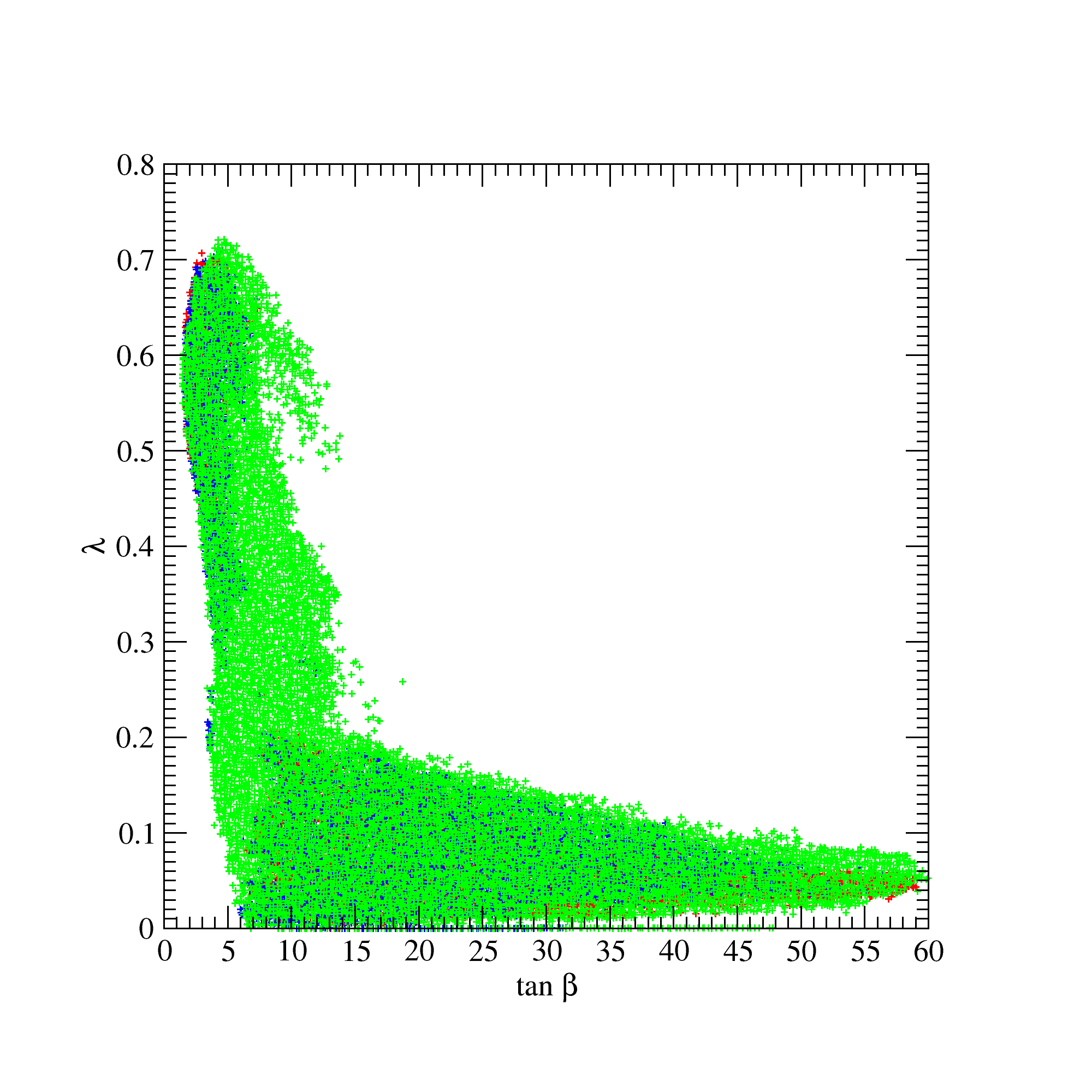}
\hspace*{-2em}
\includegraphics[width=.5\textwidth]{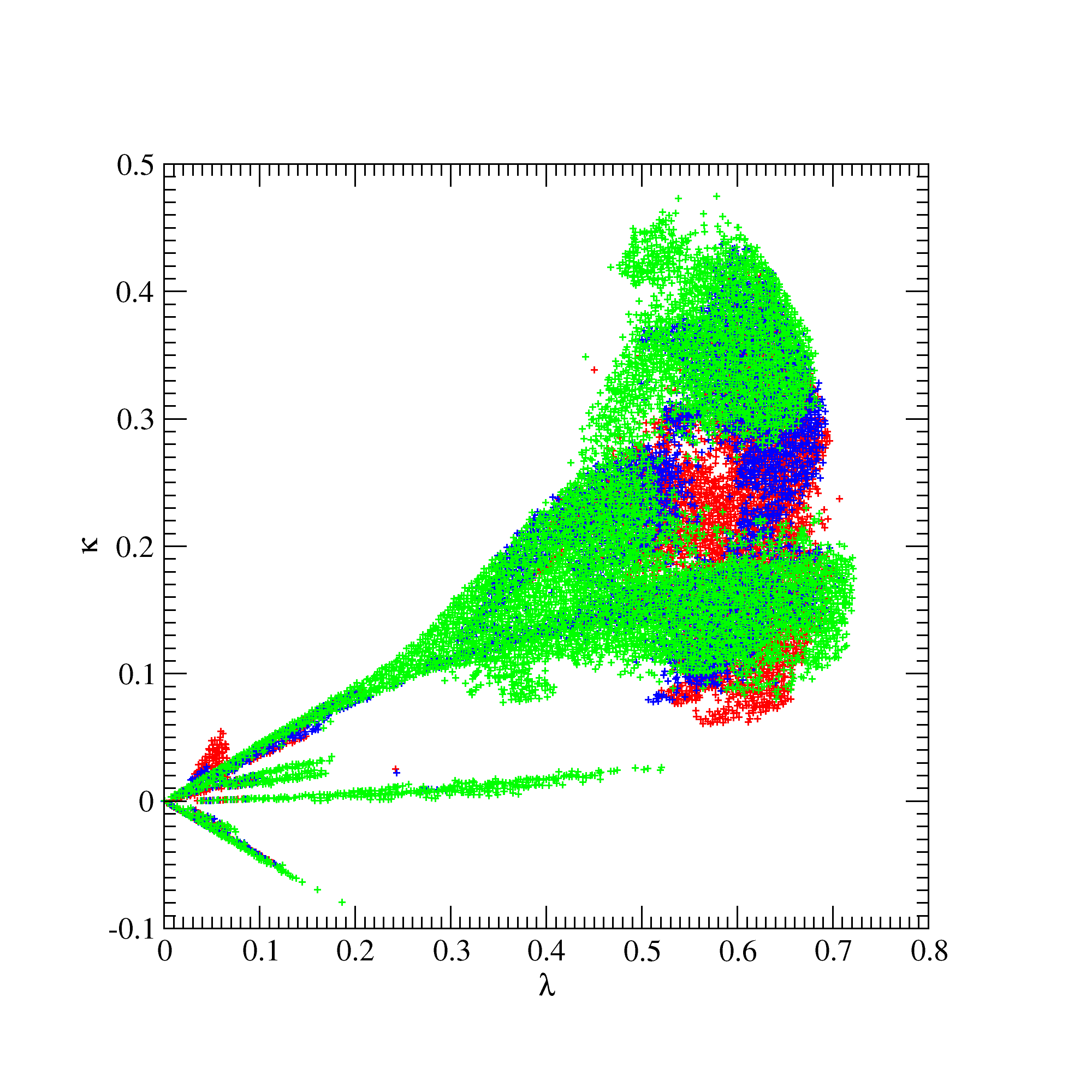}
\vspace*{-2em}
\caption{Left: $\lambda$ as function of $\tan\beta$. Right: $\kappa$ as
function of $\lambda$. The colors are as in Fig.~\ref{fig:1}.}
\label{fig:4}
\end{figure}

Since the increase of the SM-like Higgs mass with the help of the tree
level contribution $\sim \lambda^2 \sin^2 2\beta$ is effective only for
large $\lambda$ but relatively low $\tan\beta$, these regions are
typically correlated which is clarified on the left hand side of 
Fig.~\ref{fig:4}. On the right hand side we show the correlations
between $\lambda$ and $\kappa$ which shows that larger $\kappa$ are
typically related to larger~$\lambda$.

Herewith we conclude the discussion of the impact of LHC constraints on
$FT$ and the corresponding correlations with other parameters.

\section{Properties of dark matter}

Besides the enlarged Higgs sector, the enlarged neutralino sector of the
NMSSM can have a significant phenomenological impact. The LSP
(the lightest neutralino $\chi_1^0$) can have a dominant singlino
component and still be an acceptable candidate for dark matter. Its
relic density can be reduced to fit in the WMAP/Planck window, amongst
others, via the exchange of NMSSM-specific CP-even or CP-odd Higgs
scalars in the s-channel~\cite{Belanger:2005kh, Hugonie:2007vd,
Belanger:2008nt, Vasquez:2012hn}, whereas its direct detection cross
section can be very small.

\begin{figure}[ht!]\centering
\includegraphics[width=.7\textwidth]{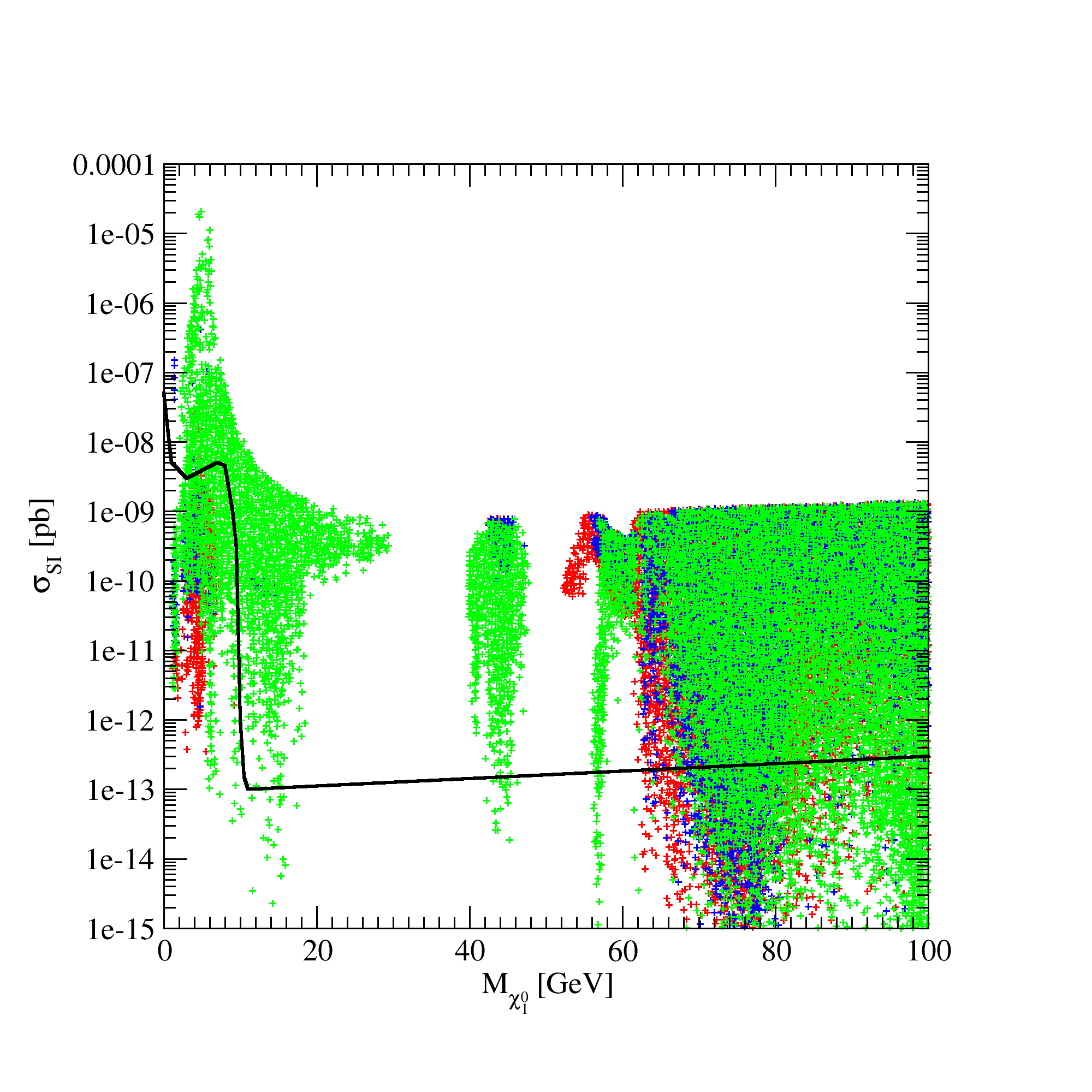}
\vspace*{-2em}
\caption{The spin-independent $\chi_1^0$-nucleon cross section
$\sigma_{\mathrm SI}$ (after
imposing constraints from the LUX experiment~\cite{Akerib:2013tjd}) as
function of $M_{\chi_1^0}$, focussing on $M_{\chi_1^0}<100$~GeV. The
black line indicates the expected neutrino background to future direct
dark matter detection experiments (from~\cite{Billard:2013qya}).
The colors are as in Fig.~\ref{fig:1}.}
\label{fig:5}
\end{figure}

The latter feature is clarified in Fig.~\ref{fig:5} where we show
the spin-independent $\chi_1^0$-nucleon cross section (after imposing
constraints from the LUX experiment~\cite{Akerib:2013tjd}) as function
of $M_{\chi_1^0}$. We
focus on $\chi_1^0$ masses below 100~GeV since no additional interesting
features appear for larger $M_{\chi_1^0}$, but the region of small
$M_{\chi_1^0}$ exhibits structures which ask for explanations.

In Fig.~\ref{fig:5} we have indicated the expected neutrino background
to future direct dark matter detection experiments from~\cite{Billard:2013qya}
as a black line; it will be difficult to
impossible to measure $\chi_1^0$-nucleon cross section smaller than this
background. Unfortunately we see that significant regions in the
NUH-NMSSM parameter space -- notably for $M_{\chi_1^0} \lsim 10$~GeV or
$M_{\chi_1^0} \gsim 60$~GeV -- may lead to such small cross sections.

\begin{figure}[ht!]\centering
\includegraphics[width=.7\textwidth]{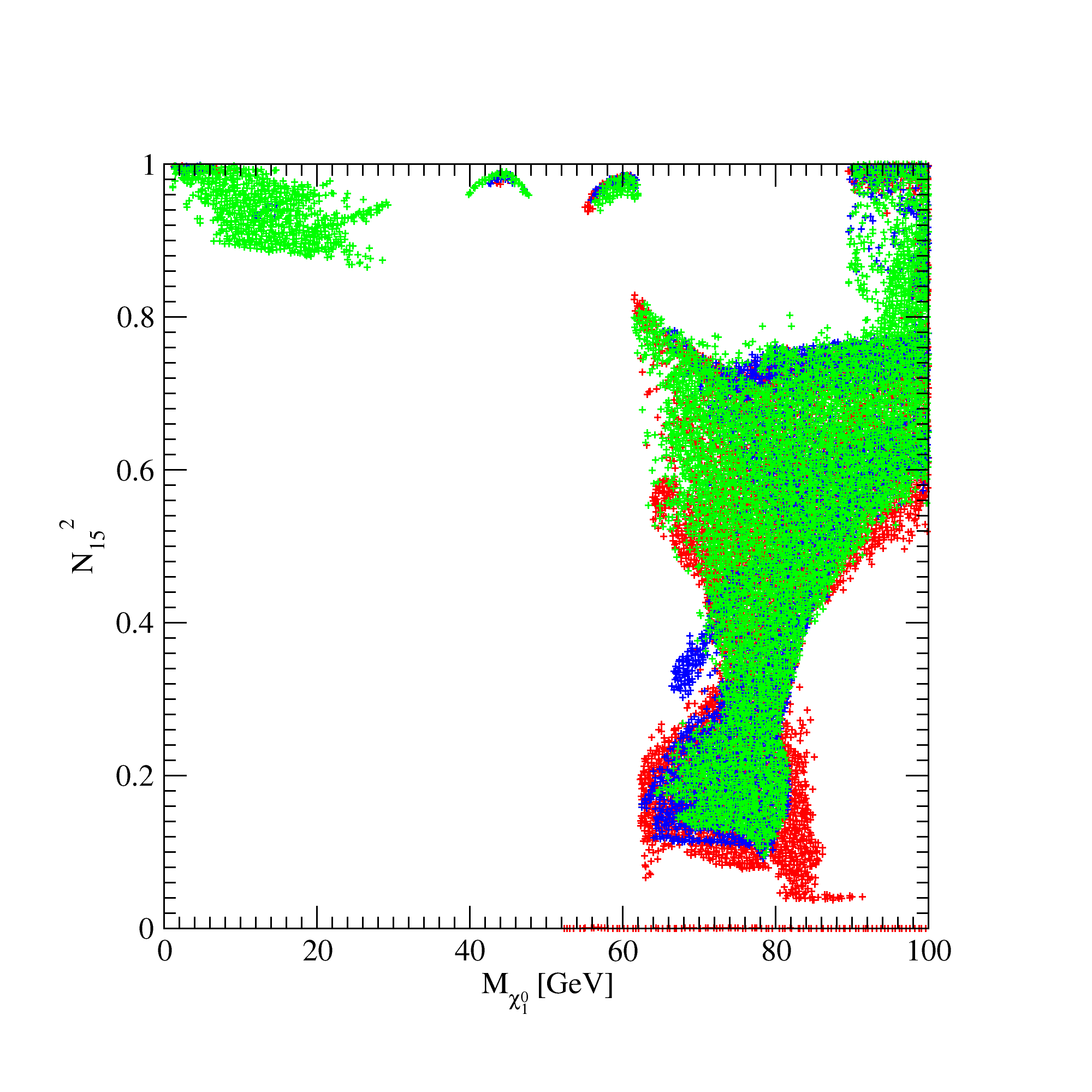}
\vspace*{-2em}
\caption{The $\chi_1^0$ singlino component (squared) as function of
$M_{\chi_1^0}$.  The colors are as in Fig.~\ref{fig:1}.}
\label{fig:6}
\end{figure}

Small $\chi_1^0$-nucleon cross sections originate from a large singlino
component of $\chi_1^0$. Its singlino component $N_{15}^2$ is shown as
function of $M_{\chi_1^0}$ in Fig.~\ref{fig:6}. 

Different regions of $M_{\chi_1^0}$ correspond to different dominant
diagrams contributing to $\chi_1^0-\chi_1^0$ annihilation before its
freeze-out. For small $M_{\chi_1^0} \lsim 30$~GeV these are the exchange
of NMSSM-specific CP-even or CP-odd Higgs scalars with masses $\approx 2
M_{\chi_1^0}$ in the s-channel, with couplings originating from the
cubic $S^3$ term proportional to $\kappa$ in the superpotential. For
$M_{\chi_1^0} \sim 40-48$~GeV, $\chi_1^0-\chi_1^0$ annihilation is
dominated by $Z$-exchange in the s-channel. The larger is the singlino
component of $\chi_1^0$, the closer $M_{\chi_1^0}$ has to be to $M_Z/2$
in order to compensate for the smaller coupling. For $M_{\chi_1^0} \sim
55-62$~GeV, $\chi_1^0-\chi_1^0$ annihilation is dominated by
$H_{SM}$-exchange. In the empty regions for $M_{\chi_1^0} \lsim 55$~GeV,
the non-singlet components of $\chi_1^0$ would have to be so large for
successful $\chi_1^0-\chi_1^0$ annihilation that the $\chi_1^0$-nucleon
cross section would violate constraints from LUX. For $M_{\chi_1^0}
\gsim 62$~GeV $\chi_1^0$ can have sizeable bino and/or higgsino
components allowing for numerous additional (e.g. MSSM-like)
$\chi_1^0-\chi_1^0$ annihilation channels.

\section{Properties of the lighter Higgs boson $H_1$}

In this paper we focus on scenarios where mixing of the SM-like Higgs
boson $H_{SM}$ with a lighter NMSSM-specific mostly singlet-like Higgs
boson $H_1$ helps to increase the mass of $H_{SM}$. This is possible
even for relatively small values of $\lambda \approx 0.1$ and moderate
to large values of $\tan\beta$~\cite{Badziak:2013bda}.

However, the $H_{SM}-H_1$ mixing angle must not be too large: It leads
to a reduction of the $H_{SM}$ couplings to electroweak gauge bosons and
quarks, hence to a reduction of its production cross section at the LHC.
These must comply with the measured signal rates, for which we require
values inside the 95\% CL contours of~\cite{Belanger:2013xza}. Moreover,
for $M_{H_1} \lsim 114$~GeV, $H_1$ must satisfy constraints from Higgs
searches at LEP~\cite{Schael:2006cr}. 

Hence the question is whether there are realistic prospects for the
discovery of $H_1$ at the LHC~\cite{Cacciapaglia:2013ora}. First we
consider the case where $H_1$ does not decay dominatly into pairs of
lighter NMSSM-specific CP-odd Higgs bosons. The branching fractions of
$H_1$ into $ZZ$ and $W^+ W^-$ are small, both due to its smaller mass
and its reduced couplings to $ZZ$ and $W^+ W^-$.

The branching fraction of $H_1$ into $\gamma\gamma$ can be
considerably larger than the one of a SM-like Higgs boson of the same
mass~\cite{Ellwanger:2010nf,Badziak:2013bda}, both due to a possible
reduction of its width into the dominant $b\bar{b}$ channel through
mixing, and/or due to additional (higgsino-like) chargino loops
contributing to the $H_1-\gamma\gamma$ coupling where the latter involve
the NMSSM-specific coupling $\lambda$~\cite{SchmidtHoberg:2012yy,
Choi:2012he}.

\begin{figure}[ht!]\centering
\includegraphics[width=.7\textwidth]{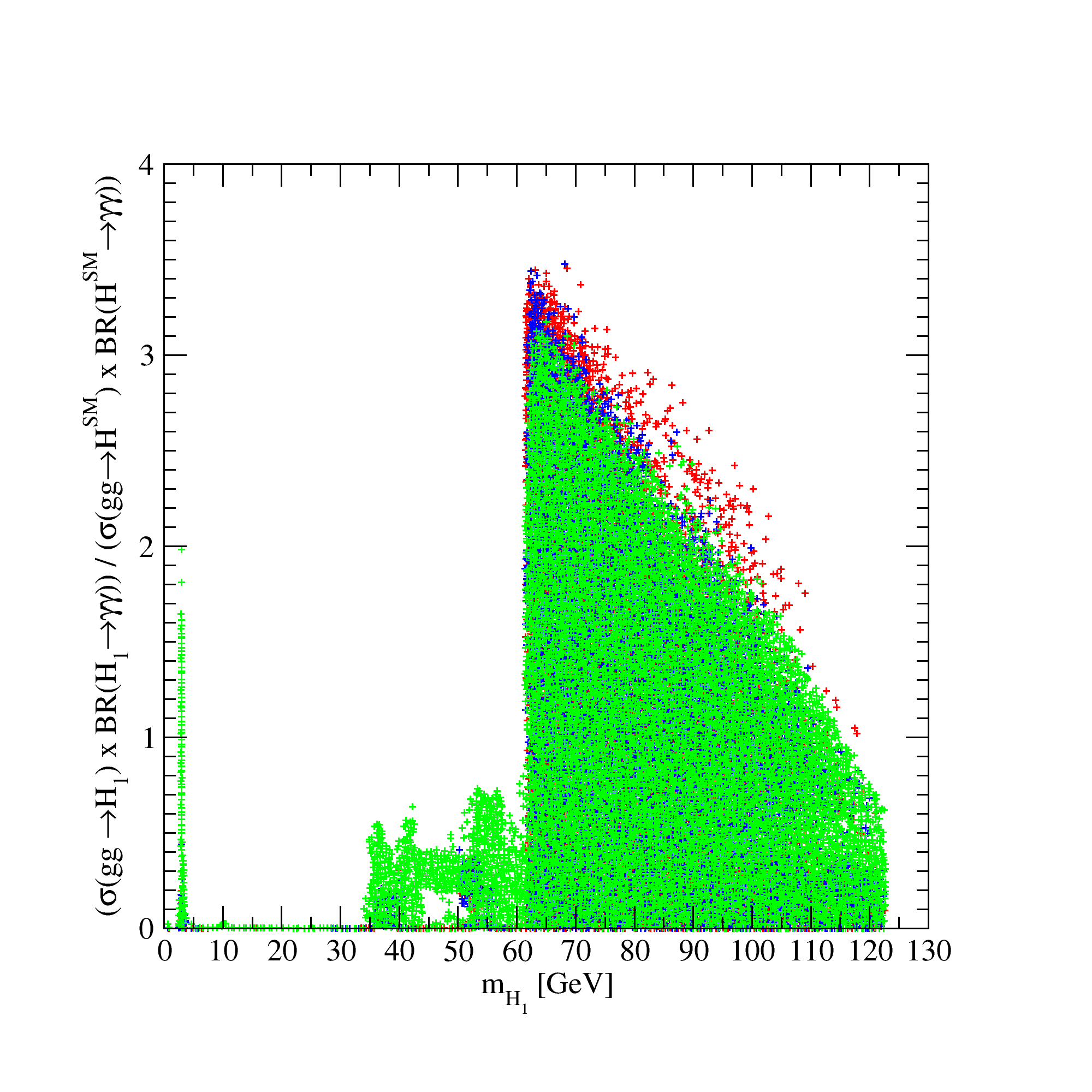}
\vspace*{-2em}
\caption{The $H_1$ signal rate in gluon fusion and the $\gamma\gamma$
channel relative to a SM-like Higgs boson $H^{SM}$ of the same
mass. The color code is as in Fig.~\ref{fig:1}.}
\label{fig:7}
\end{figure}

However, due to the reduced coupling of $H_1$ to SM particles, its
production cross section $\sigma_{H_1}$ is smaller than the one of a
SM Higgs boson $H^{SM}$ of the same mass. Hence one has to consider
the reduced signal rate $\sigma_{H_1}\times BR(H_1\to\gamma\gamma)/
\left(\sigma_{H^{SM}}\times BR(H^{SM}\to\gamma\gamma
)\right)$~\cite{Ellwanger:2010nf, Cao:2011pg,Badziak:2013bda,King:2012tr}
which is shown for production via gluon fusion in Fig.~\ref{fig:7}.

We see that the signal rate can be about 3.5 times larger than the one
of a SM-like Higgs boson of a mass of $\sim 60$~GeV. The absence of
points with large signal rates for $M_{H_1} \lsim 60$~GeV follows from
the constraints on the signal rates of $H_{SM}$: For $M_{H_1} \lsim
60$~GeV, $H_{SM}$ could decay into a pair of $H_1$ bosons, and this
decay channel is easily dominant if kinematically allowed. The
corresponding reductions of the other $H_{SM}$ branching fractions would
be incompatible with its measured signal rates. (The possible
enhancement of the signal rate for $M_{H_1}\lsim 3.5$~GeV originates
from the absence of decays into $b\bar{b}$ and $\tau^+\tau^-$, which
makes it very sensitive to relative enhancements of the width into
$\gamma\gamma$ via chargino loops.) For $M_{H_1} \gsim 110$~GeV, some
points with a reduced signal rate $\gsim 0.5$ could actually already be
excluded by limits from CMS in~\cite{CMSgamgam} depending, however, on
the relative contribution of gluon fusion to the expected signal rate in
this mass range. On the other hand it is clear that, for $M_{H_1} \sim
M_Z$, the $H_1\to\gamma\gamma$ channel faces potentially large
backgrounds from fake photons from $Z\to e^+ e^-$~decays.

For the $b\bar{b}$ and $\tau^+\tau^-$ final states we found that due to
the reduction of the production cross section and the reduction of the
couplings (i.e. branching fractions) of $H_1$ its reduced signal rates
in gluon fusion, vector boson fusion and associate production with $Z/W$
are always below 0.3 for $M_{H_1} \lsim 114$~GeV, and still below 0.6
for $114\ \mathrm{GeV}\ \lsim M_{H_1} \lsim 126$~GeV; hence we will not
further analyse these channels (also plagued by the absence of narrow
peaks in the invariant mass of the final states).

\begin{figure}[ht!]\centering
\includegraphics[width=.5\textwidth]{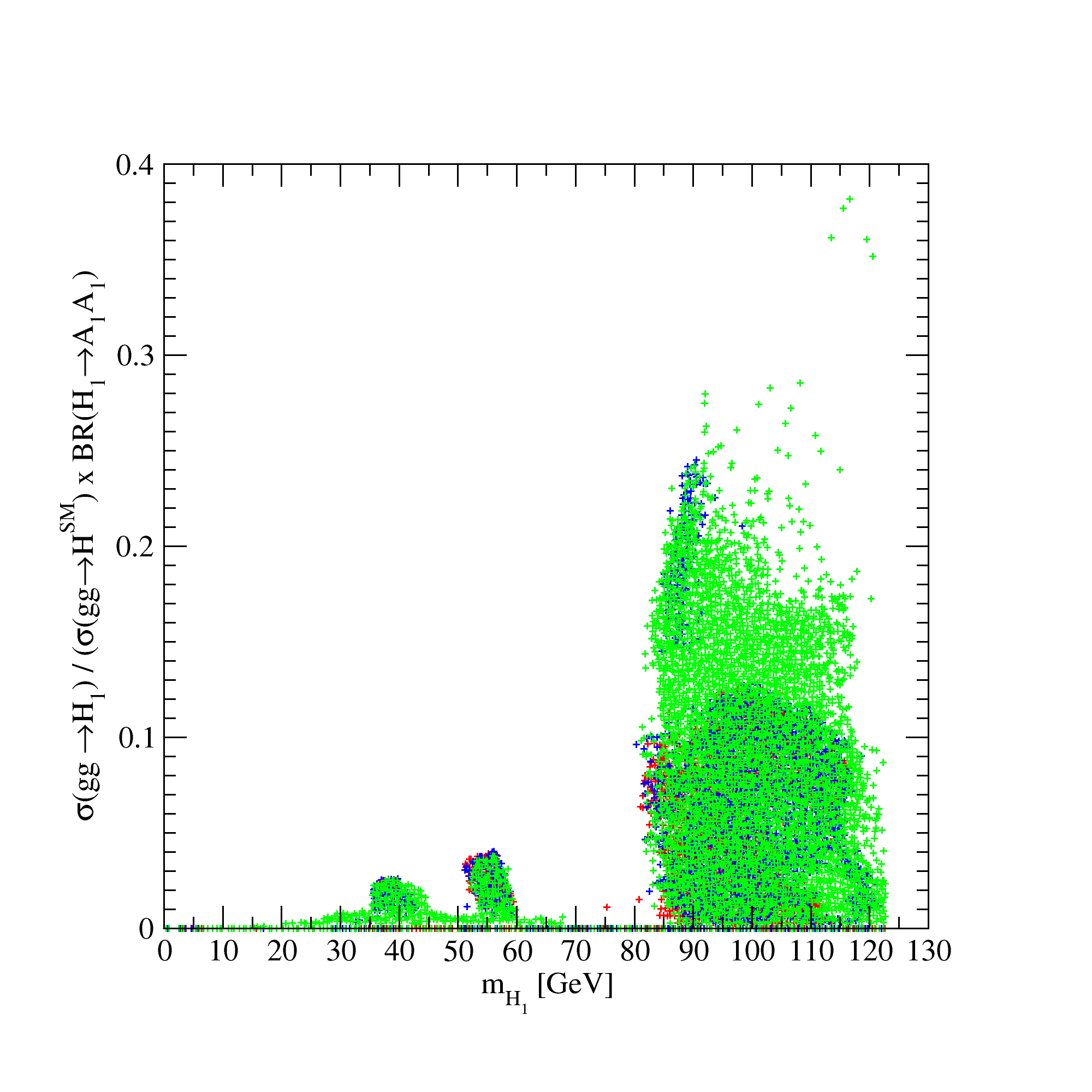}
\hspace*{-2em}
\includegraphics[width=.5\textwidth]{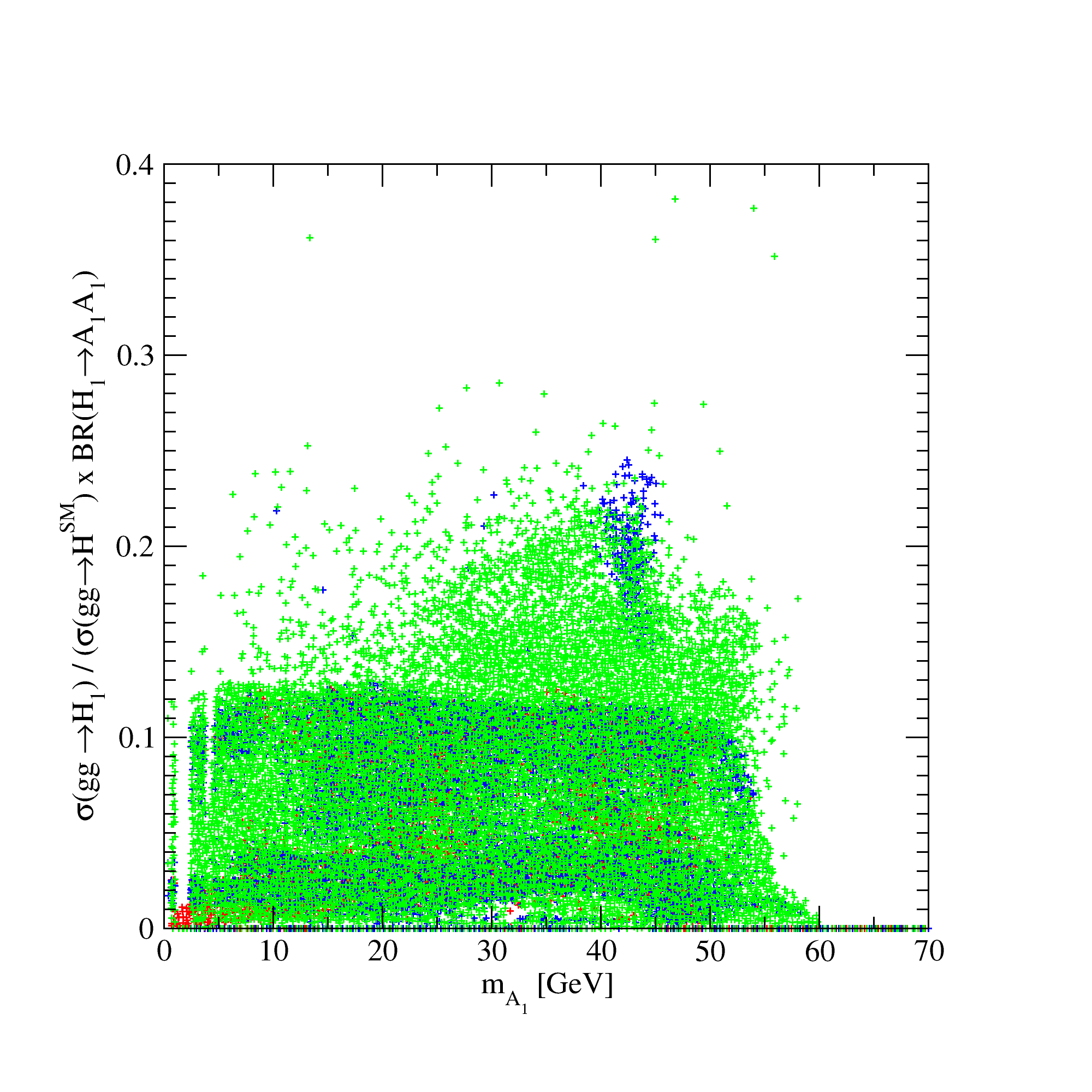}
\vspace*{-2em}
\caption{$\sigma_{H_1}(ggF)/\sigma_{H^{SM}}(ggF)\times BR(H_1\to A_1
A_1)$ as function of $M_{H_1}$ (left) and $M_{A_1}$ (right). The color
coding is as in Fig. \ref{fig:1}.}
\label{fig:8}
\end{figure}

Another possibility is that $H_1$ decays dominantly into pairs of light
NMSSM-specific CP-odd Higgs bosons $A_1$ (see~\cite{Cerdeno:2013cz} and
refs. therein). If this channel is open, the corresponding branching
fraction $BR(H_1 \to A_1 A_1)$ can vary from 0 to 1 for all $M_{H_1}$
and $M_{A_1}$. However, the production cross section of $H_1$ is always
reduced relative to the one of a SM-like Higgs boson $H^{SM}$ of the
same mass. Focussing again on gluon fusion, we show in Figs.~\ref{fig:8}
the $BR(H_1 \to A_1 A_1)$ multiplied by the reduced $H_1$ production
cross section (relative to the one of a SM-like Higgs boson $H^{SM}$ of
the same mass) as function of $M_{H_1}$ and $M_{A_1}$.

The dominant decay branching fractions of $A_1$ are very similar to the
ones of a SM-like Higgs boson of the same mass, i.e. dominantly into
$b\bar{b}$ and $\tau^+\tau^-$ if kinematically allowed. These
unconventional channels $H \to A_1 A_1 \to ...$ have been searched for
at LEP by OPAL~\cite{Abbiendi:2002qp,Abbiendi:2004ww,Abbiendi:2002in},
DELPHI~\cite{Abdallah:2004wy} and ALEPH~\cite{Schael:2010aw}. The
corresponding constraints are taken into account in {\sf NMSSMTools},
and explain the absence of sizeable signal rates for $M_{H_1} \lsim
80$~GeV. For $M_{H_1} \gsim 86$~GeV and, simultaneously, $0.25~
\mathrm{GeV} \lsim
M_{A_1} \lsim 3.55$~GeV, first LHC analyses by CMS~\cite{Chatrchyan:2012cg}
have lead to upper limits on the signal cross
section for $H \to A_1 A_1 \to 4\mu$ which exclude some of the points in
this range of $M_{A_1}$.

For heavier $A_1$ leading to dominant $b\bar{b}$ and/or $\tau^+\tau^-$
decays, analyses of possible signals are certainly more difficult. At
least we find that, for $M_{H_1} \gsim 80$~GeV, production cross
sections times branching fractions can be relatively large without
violating present constraints, which should motivate future analyses of
these channels.

Concerning the signal rates of the SM-like Higgs boson $H_2$ we
remark that all values allowed by the 95\% confidence level contours
in~\cite{Belanger:2013xza} in the planes of Higgs production via
(gluon fusion and ttH) -- (vector boson fusion and associate
production with W/Z) for Higgs
decays into $\gamma\gamma$, ZZ+WW and $b\bar{b}+\tau^+\tau^-$ have been
found by our scan.

Also possible are decays of $H_2$ into pairs of light CP-even or CP-odd
states $H_1$ or $A_1$. They are limited by the SM-like signal rates of
$H_2$, but branching fractions of up to 40\% are still allowed.

\section{Conclusions and outlook}

In spite of the recent constraints on the mass and the signal rates at
the LHC on a SM-like Higgs boson, upper bounds on signal rates generated
by first generation squarks and gluinos and upper bounds on dark matter
-- nucleus cross sections we have seen that large ranges of the
parameter space of the NUH-NMSSM remain viable. Within this scenario,
bounds from squark/gluino searches dominate the lower bounds on
fine-tuning which remain, on the other hand, considerably smaller than
in the (NUHM-)MSSM and more constrained versions of the NMSSM.

The mass of the LSP is barely constrained, up to some ``holes'' around 30
and 50 GeV, and can possibly be below 1~GeV. Due to its possibly
dominant singlino component, its direct detection cross section can be
considerably smaller than the neutrino background, which makes it
compatible with all future null-results in direct (and actually also
indirect) dark matter searches.

We have not discussed all possible NUH-NMSSM-specific phenomena at
colliders, which would be beyond the scope of the present paper. Here we
focussed on the properties of an additional lighter NMSSM-specific Higgs
boson $H_1$, in particular on its signal rates in channels which are
accessible at the LHC. These include the potentially promising diphoton
decay channel, but also $H_1$-decays into a pair of even lighter CP-odd
bosons $A_1$. Albeit taking into account all present constraints on
additional lighter Higgs bosons, wide ranges of $H_1$ and $A_1$ masses
remain to be explored.

Amongst additional NUH-NMSSM-specific phenomena at colliders -- induced
by a sing\-lino-like LSP and/or additional Higgs states -- are possibly
unconventional cascade decays of charginos and top- and bottom-squarks,
which require additional studies. Future work will also be dedicated to
the possibilities for and signatures of Higgs-to-Higgs decay cascades
induced by heavier Higgs states in the NUH-NMSSM.

\section*{Acknowledgements}

UE acknowledges support from the ERC advanced grant Higgs@LHC and from
the European Union Initial Training Networks INVISIBLES
(PITN-GA-2011-289442) and  Higgs\-Tools (PITN-GA-2012-316704).
The authors acknowledge the support of France Grilles for providing
computing resources on the French National Grid Infrastructure.

\end{document}